\documentstyle[11pt,aaspp4,flushrt]{article}

\def\pc{\,{\rm pc}}
\def\kpc{\,{\rm kpc}}
\def\Mpc{\,{\rm Mpc}}

\def\eV{{\,\rm eV}}

\def\cmm2{{\,\rm cm^{-2}}}
\def\cm2{{\,{\rm cm}^2}}
\def\cmm3{{\,{\rm cm}^{-3}}}
\def\gcmm3{{\,{\rm g\,cm^{-3}}}}

\def\g{{$\gamma$}}
\def\G{\,{\rm G}}
\received{August 15, 1995}
\revised{REVISION DATE}
\accepted{ACCEPT DATE}
\cpright{AAS}{1995}

\journalid{VOL}{JOURNAL DATE}
\articleid{START PAGE}{END PAGE}
\paperid{MANUSCRIPT ID}
\cpright{AAS}{1995}
\ccc{CODE}

\slugcomment{FERMILAB-Pub-95/282-A}
\lefthead{Lee, Olinto and Sigl}
\righthead{Magnetic Field and Cosmic Rays}

\begin{document}
\title{EXTRAGALACTIC MAGNETIC FIELD AND THE HIGHEST ENERGY COSMIC
RAYS\footnote{Submitted to \it The Astrophysical Journal Letters}}
\author{Sangjin Lee$^{1,2}$, Angela V.~Olinto$^{1,2}$, \& G\"{u}nter
Sigl$^{1,2}$}
\affil{$^1$Department of Astronomy \& Astrophysics\\
Enrico Fermi Institute, The University of Chicago, Chicago, IL 60637-1433\\
$^2$NASA/Fermilab Astrophysics Center\\
Fermi National Accelerator Laboratory, Batavia, IL 60510-0500}

\begin{abstract}
The strength and spectrum of the extragalactic magnetic field are
still unknown. Its measurement would help answer the question of whether
galactic fields are  purely a primordial relic or were
dynamically enhanced from a much smaller cosmological seed field. In
this letter, we show that the composition, spectrum, and directional
distribution of extragalactic ultrahigh energy cosmic rays with
energies above $\simeq10^{18}\eV$ can probe the large scale
component of the extragalactic magnetic field below the present
observational upper limit of $10^{-9}$ Gauss.
Cosmic ray detectors under construction or currently in the proposal
stage should be able to test the existence of the extragalactic magnetic
fields on scales of a few to tens
of Mpc and strengths in the range $\simeq 10^{-10}-10^{-9}$ Gauss.
\end{abstract}

\section{Introduction}

Magnetic fields are   universally present in astronomical
bodies ranging from the Earth to the distant quasars, but it is still unknown
if
magnetic fields permeate the universe as a whole. Astrophysical magnetic
fields
may arise due to the existence of a primordial magnetic field that grows as
galaxies form.  The discovery of an extragalactic magnetic
field on scales larger than virialized systems (i.e., larger than clusters
of galaxies) would reveal the presence of a primordial field. The existence
of such a primordial field would  help understand the origin of astrophysical
magnetic
fields and may open a new window into processes ocurring in the early
universe.

In this letter, we show that the study of ultra
high energy cosmic rays (UHE CRs) with energies above
$\simeq10^{18}\eV$ can probe primordial fields below the current
upper bound. Our proposed method is complementary to traditional ones
(e.g., Kronberg 1994) and more recent suggestions (e.g., Plaga
1995). At present,  the most stringent  constraint on large scale extragalactic
fields comes from limits on the  Faraday rotation of light coming to us from
distant quasars. The upper bound on a widespread, all-pervading field
is $\sim 10^{-9}$ Gauss (e.g., Kronberg 1994). There are weaker constraints
derived from the synchrotron emission from nearby galaxy
clusters (Kim et al.~1989)
and the cosmic microwave background isotropy.

UHE CR nucleons from extragalactic sources
are attenuated
in energy while propagating through the cosmic microwave background
(CMB). Above a few $10^{19}\eV$, nucleons produce
pions on the CMB photons and the energy of the cosmic ray nucleons is degraded
rapidly, which is known as the Greisen-Zatsepin-Kuz'min (GZK)
effect (Greisen 1966;
Zatsepin \& Kuz'min 1966). Produced pions, on the other hand, decay into
secondary leptons such as electrons, muons, and neutrinos, and photons. Since
muons decay to electrons and neutrinos, the final secondary particles are
photons, electrons, and neutrinos, among which photons are more readily
detectable.

Very energetic secondary photons and electrons couple to form an
electromagnetic (EM) cascade. The $\gamma$-rays produce electron
pairs on the CMB and the universal radio background photons
(see Lee \& Sigl 1995, Lee 1995 and references therein for more
detailed discussions). The resulting electrons (or
positrons) in turn upscatter background photons via inverse Compton
scattering (ICS), thus completing a cycle. It is through these two processes
that an
EM cascade develops in the intergalactic medium. As a result, if one
propagates a purely protonic spectrum, one gets a processed nucleon
spectrum with the GZK cutoff, and a secondary EM (photons and
electrons) spectrum.

The extragalactic magnetic field (EGMF) influences the UHE CR flux
mainly through their charged components, namely the primary hadrons
and the secondary electrons. In the energy range under
consideration the hadrons are deflected with negligible
energy loss due to synchrotron radiation whereas the electrons are
negligibly deflected before they lose most of their energy.
In \S 2, we explore the effect of the electron energy loss due to the
EGMF on the secondary EM cascade spectrum. We then  discuss the deflection
of the hadronic component in the EGMF, in \S 3. Finally  in \S 4, we
summarize our findings.

\section{Extragalactic Magnetic Field and Ultrahigh Energy \g-Ray Flux}
The EGMF plays a crucial role in the development of
the EM cascade. In the presence of the magnetic field, electrons lose energy
via synchrotron radiation loss, which is given by
\begin{equation}
\frac{dE}{dt} = - \frac{e^4 B^2}{24\pi^2 m_e^4} E^2 = -\frac{2}{3} r_0^2 B^2
\left( \frac{E}{m_e} \right)^2,
\end{equation}
where $B$ is the strength of the large scale EGMF, $r_0$ is the classical
electron radius, and $m_e$ is the electron mass.

In  fig.~1, we show the rates of ICS and synchrotron loss for
electrons. Whereas the rate of ICS responsible for EM cascade
development decreases with energy, the synchrotron loss rate
increases with energy. Therefore, in a narrow energy range a transition occurs
between a regime where ICS is dominant and electrons couple to
photons efficiently and another where electrons are rapidly lost due to
synchrotron loss and the cascade is suppressed.
Below $\simeq 10^{20}\eV$ (the threshold for pair production on the
radio background), cascade development in the absence of an EGMF
would give rise to  a generic power law photon spectrum with index
$\simeq-1.5$. The above mentioned transition in the secondary
\g-ray spectrum will therefore occur between this generic
cascade shape and a synchrotron loss dominated spectrum.
As long as synchrotron quanta can be neglected the latter is
given by the photons produced ``directly'' by source injection
or from pion production by nucleons, before undergoing pair
production in the low energy photon background.

In a magnetic field of strength $B$ measured in G (Gauss) the
synchrotron spectrum produced by an electron of energy $E_e$ peaks at
\begin{equation}
E_{\rm syn} \simeq 6.8 \times 10^{13} \left( \frac{E_e}{10^{21} \eV} \right)^2
\left( \frac{B}{10^{-9}\G} \right) \eV\,,\label{synch}
\end{equation}
and falls off exponentially at higher energies. In the following
we assume that the observable EM flux is energetically dominated
by \g-rays and electrons with energy $E\la10^{21}\eV$. Then,
according to eq.~[\ref{synch}], the contribution of synchrotron
radiation to the \g-ray flux above $10^{18}\eV$ can be safely
neglected as long as $B\la10^{-6}\G$ everywhere.

The energy where the transition in the \g-ray spectrum occurs is
in general a function of the magnetic field
strength and the background photon spectrum, but {\em not\/} a function of the
source  distance or the injection spectrum.
If we consider only the CMB, the relation between the transition
energy $E_{\rm tr}$ and the magnetic field strength is
$E_{\rm tr}\propto B^{-1}$. In order to include the less well
known diffuse extragalactic radio background into consideration we
adopt its usual description by a power law with an overall amplitude and a
lower frequency cutoff as parameters (Clark, Brown, \& Alexander 1970)
the latter one being
the main source of uncertainty. Using a cutoff at 2 MHz as
suggested by Clark et al.~(1970) the above relation is modified to
\begin{equation}
E_{\rm tr} \simeq 10^{19}
\left(\frac{B}{10^{-9}\G}\right)^{-1.3}~\eV ~~(B \ga 10^{-10}\G) \ .
\end{equation}
For the same magnetic field
a cutoff at lower frequencies would increase the rate of ICS of
the then more abundant low frequency radio photons and
thus the value for $E_{\rm tr}$ (see fig.~1). Assuming that the
radio cutoff frequency lies somewhere in the range between 0.5
MHz and 3 MHz, for a given $E_{\rm tr}$ the EGMF strength $B$ is
uncertain within about a factor 5.

Therefore, for $B\ga10^{-10}\G$ it is possible to approximately
determine the EGMF strength by
searching for a dip in
the $\gamma$-ray flux below $10^{20}\eV$ which would mark a
transition between an ICS and a synchrotron loss dominated
regime. For $B \la 10^{-10}\G$, this transition occurs above the
pair production threshold on the radio background where
the $\gamma$-ray flux increasingly
depends on  several unknown factors such as the charged
cosmic ray flux above the GZK cutoff.
Thus, even though the $\gamma$-ray flux can be comparable
to the nucleon flux above $10^{20}\eV$,
a discussion of possible magnetic field signatures in its
spectrum would presently be too speculative.

We developed a numerical code for the propagation of
nucleons, photons, and electrons through the intergalactic medium
which employs a transport equation formalism, the details of which can
be found in Lee (1995).
The observed UHE CR flux below $10^{20}\eV$ (see, e.g., Bird et
al.~1994; Yoshida et al.~1995) is reproduced quite well by a
diffuse distribution of sources injecting protons with a
spectrum $\propto E^{-2.3}$ up to some maximal energy
considerably beyond the GZK cutoff (Yoshida \& Teshima 1993; Sigl
et al.~1995). For the calculations presented here
we therefore adopted this proton injection
spectrum with a maximal energy of $10^{22}\eV$ (see figures).
The EGMF enters the calculation via the
synchrotron loss of electrons.

The transition between ICS and synchrotron loss domination can
be easily seen in fig.~2, which shows the processed nucleon and photon
spectra for a single source at a distance of 30 Mpc for a range
of EGMF strengths. In general, one expects a distribution of
cosmic ray sources rather than a single source
at a fixed distance. In fig.~3, we show the diffuse spectrum
from a continuous source distribution extending
up to 1 Gpc. We assume a flat universe with zero cosmological constant and a
Hubble constant of $H_0 = 75$ km sec$^{-1}$Mpc$^{-1}$, and a comoving source
density scaling as $(1+z)^2$ in
redshift $z$ as in some ``bright phase'' models of CR sources
(e.g., Yoshida \& Teshima 1993; Hill \& Schramm 1985 and
references therein). The results are not very sensitive to these choices.
In the diffuse case, the \g-ray to nucleon flux ratio tends to be
smaller than for a single source, and the
spectral features are not as pronounced, but still
detectable for $B \ga10^{-10}\G$.

The ``extragalactic magnetic field'' in this analysis
refers to the average  component of the EGMF normal to the
line of propagation. Primordial magnetic fields are expected to have very
little
structure on scales below $\sim $ few Mpc (Jedamzik, Katalinic, \& Olinto
1995),
but condensed structures such as galaxies and clusters of galaxies can
``polute"
the intergalactic medium with stronger magnetic fields on smaller scales.
Fortunately,  the effect of the EGMF on the $\gamma$-ray
spectral shape discussed here is most sensitive to the average field on large
scales. When the EM cascade goes through a strong field region, the electrons
in
the cascade lose energy rapidly and drop out of the UHE range.
In contrast, the UHE nucleon and \g-ray fluxes are usually
hardly affected directly by the radiation field of the
intervening object (Stecker et al.~1991; Szabo \& Protheroe 1994; Norman,
Melrose, \& Achterberg 1995).
After escaping the object, the cascade redevelops
and the cascade spectrum recovers quickly at only a slightly
smaller amplitude. The influence of
intervening
objects decreases with increasing strength of the large scale field and
becomes important only when the objects
are very close to the observer (e.g., less than $\sim 5\Mpc$
away), or when
their linear size is significant compared to the total
propagation distance. We can derive conditions for the filling
factors for such objects by requiring that they are much more sparsely
populated along the line of sight to the source
than the typical cascade regeneration length $s_c$ (e.g., $\sim$ 5
Mpc). For an average linear size $\bar l_{\rm iv}$ of the
intervening objects their filling factor $f_{\rm iv}$ must satisfy
\begin{equation}
f_{\rm iv} \ll \frac{\bar l_{\rm iv}}{s_c}.
\end{equation}
For field galaxies the above relation is $f_g \ll 10^{-3}$, and for clusters of
galaxies $f_c \ll 0.1$. The actual filling factors for galaxies,
$f_g \la 10^{-7}$, and galaxy clusters, $f_c \la
10^{-4}$ (Kolb \& Turner 1992; Nichol, Briel, \& Henry 1994)
satisfy these constraints. Therefore,
intervening objects do not modify the above discussion substantially. One
interesting exception may be the effect of nearby large structures such as
the Virgo Cluster. If Virgo has strong magnetic fields, e.g.~of order
$10^{-7}\G$ on Mpc scales, the UHE \g-ray flux from background sources might
be modified across Virgo's angular extension (see fig.~2).

The ability of future detectors to study EGMF features crucially
depends on the \g-ray to nucleon flux ratio. For nearby strong
sources, the secondary \g-ray flux could be measurable with
instruments which are sensitive to ratios down to $\simeq1\%$.
This could be achieved by the proposed Pierre Auger Project
(see e.g., Boratav et al.~1992), which would also allow an
angular resolution of $\simeq1^{\circ}$. The case of a diffuse source
distribution is more challenging; the $\gamma$-ray flux is typically smaller
and the EGMF feature is less pronounced than for a single source
at moderate distances (see figs.~2 \& 3), but the dip in the
$\gamma$ spectrum may still be detectable. Thus, measuring the
\g-ray flux between $\simeq10^{18}\eV$ and $\simeq10^{20}\eV$
has the potential to either detect or find strong evidence against an EGMF
$\ga 10^{-10}\G$.

\section{Charged Cosmic Ray Deflection by Extragalactic Magnetic
Fields}
Here, we discuss the influence of the EGMF
on the charged UHE CR flux from discrete sources.  We
restrict ourselves to the case of small deflection angles (for
the opposite limit see, e.g. Wdowczyk \& Wolfendale 1979;
Berezinskii, Grigor'eva, \& Dogel' 1989).
In this case, the energy spectrum of charged UHE CRs from a given
source is not significantly altered as compared to a
straight-line propagation. However, if the sources are strong
enough to cause an anisotropy in the UHE CR flux, the
directional correlation of ``hot spots" with
possible  sources will depend on the  EGMF. The following discussion relates
to this anisotropic component of the charged UHE CR flux.

As in \S 2, let us assume that the large
scale EGMF can be characterized by a typical field strength $B$
and a coherence length $l_c$. Furthermore, we assume for the moment
that the source distance $r$ is smaller than the energy attenuation
length $\lambda=E(dE/dr)^{-1}$ for a charged cosmic ray of energy $E$
which can then be treated as approximately constant throughout
propagation. For nucleons, $\lambda\simeq10$ Mpc above the GZK cutoff
(at $E\simeq6\times10^{19}\eV$), and $\lambda\simeq1$ Gpc much
below the GZK cutoff. A more sophisticated analysis would require
a Monte Carlo simulation of both UHE CR propagation and deflection.
However, since data on both UHE CR and the EGMF are so sparse to date,
we feel that a qualitative discussion of the principle effects
is more appropriate at the moment. We now consider two cases.

(i) The source distance is smaller than the coherence length,
$r\la l_c$. Then, in vectorial notation, the deflection
angle ${\bf\alpha}$ is given by
\begin{equation}
  {\bf\alpha}=-{Ze\over E}\,{\bf r}\times{\bf B}=
  5.3^{\circ}\,Z\left({E\over10^{20}\eV}\right)^{-1}\left({r\over10\Mpc}
  \right)\left({B\over10^{-9}\G}\right)\,
  \left({\bf\hat r}\times{\bf\hat B}\right)\,,\label{def1}
\end{equation}
where ${\bf r}$ is the radius vector pointing to the source, $Z$
is the charge of the UHE CR component, ${\bf\hat r}={\bf r}/
\vert{\bf r}\vert$, and ${\bf\hat B}={\bf B}/\vert{\bf B}\vert$.
Thus, a correlation between the UHE CR flux of charge $Z$ and energy
$E$ and source counterparts, systematically shifted by an angle
${\bf\alpha}$, would indicate that $l_c\ga r$ and for a known
source distance $r$ would allow to measure the combination
$B({\bf\hat r}\times{\bf\hat B})$. The characteristic $E$- and
$Z$ dependence of ${\bf\alpha}$ would provide an additional test
for the hypothesis that the deflection is caused by an EGMF.

(ii)  The source distance is considerably larger than the
coherence length, $r\gg l_c$. In this case the deflection angle
undergoes a diffusion process during propagation and the source
shape in the UHE CR flux will be smeared out over a typical angle
\begin{equation}
  \alpha_{\em rms}\simeq{2\over\pi}{ZeB\over E}\left(rl_c\right)^{1/2}=
  1.1^{\circ}\,Z\left({E\over10^{20}\eV}\right)^{-1}
  \left({r\over10\Mpc}\right)^{1/2}
  \left({l_c\over1\Mpc}\right)^{1/2}
  \left({B\over10^{-9}\G}\right)\,.\label{def2}
\end{equation}
Therefore, if sources appear spread out in the UHE CR flux of
charge $Z$ and energy $E$ by a typical angle $\alpha$, this would
indicate that $l_c\la r$ and for a known source distance $r$ would
allow to measure the combination $Bl_c^{1/2}$.

If the source distance is larger than the energy attenuation length,
$r\ga\lambda$, eqs.~[\ref{def1}] and [\ref{def2}] tend to overestimate
$\alpha$.
In fact, in the limit $r\gg\lambda$ the deflection angle $\alpha$
``saturates" as a function of $r$ and for approximately energy
independent $\lambda$, $r$ has to be substituted by $\lambda$ and
$\lambda/2$ in eqs.~[\ref{def1}] and [\ref{def2}], respectively.

Secondary \g-rays produced by the interactions
of the charged UHE CRs are also expected to correlate with the
sources. Due to their continuous production they will be smeared
out over angles which are typically somewhat smaller than given
in eqs.~[\ref{def1}] and [\ref{def2}].

Sources which can act as suitable probes for the EGMF via the
effects discussed above have to obey the following conditions
apart from producing a detectable anisotropic UHE CR flux
component: Their apparent angular size should be
smaller than the deflection angle $\alpha$. The same pertains to
the apparent angular radius of a possible high magnetic field
region around the source if it can cause deflections in excess
of $\alpha$. For example, a $10^{-6}\G$ field over a scale
$\ga100\kpc$ is possible in galaxy clusters (see, e.g., Kronberg
1994) and would
completely bend around a $10^{20}\eV$ proton. However, as long
as a detectable proton flux emerges from such an object and the
above conditions are fulfilled, it could still be a suitable
probe of the EGMF. Finally, there should be no intervening
high magnetic field regions between source and observer which
could cause bending by more than $\alpha$. For example, for
$r\ga l_c$, $\lambda$, this corresponds to the condition
$l_{\rm iv}B_{\rm iv}\la\left(\lambda l_c\right)^{1/2}B$
for linear scale $l_{\rm iv}$ and strength $B_{\rm iv}$ of the
intervening field. This condition could well be satisfied
along most lines of sight
since known objects with high field regions like galaxies and
galaxy clusters have a small filling factor $f\la10^{-5}$.

In light of these conditions we believe that some of the nearby
galaxy clusters and powerful field radio galaxies could well
be suitable EGMF probes since they are expected to contribute
significantly to the UHE CR flux (Rachen, Stanev, \& Biermann
1993). The accuracy to which the EGMF bending can be determined
is limited by the (to date) unknown additional bending by the
galactic magnetic field of strength $B_g$ and scale height
$l_g$. Thus, according to eq.~[\ref{def1}], the sensitivity of
deflection measurements of the EGMF is restricted to field
parameters satisfying $Br\ga10^{-9}\G\Mpc
\left(B_g/\mu{\rm G}\right)\left(l_g/300\pc\right)$ where the fudge
factors are the parameter values usually assumed for the
galactic magnetic field.

\section{Conclusions}
We discussed how composition, spectrum, and directional
distribution of UHE CR above $\simeq10^{18}\eV$ can be used to
gain information about the large scale (a few to tens of Mpc)
EGMF. Spectral features in the
$\gamma$-ray flux are sensitive to field strengths in the range
$\simeq10^{-10}-10^{-9}\G$. In a similar range,
correlations between an anisotropic charged UHE CR flux
component and possible sources could provide independent
information on the EGMF including its polarization.
Both effects should yield consistent estimates for the EGMF
strength. Strong discrete
sources detected in UHE CRs by future instruments with an angular
resolution of $1^{\circ}$ or better and a sensitivity to $\gamma$-ray
to nucleon flux ratios of $1\%$ or smaller would provide the best
conditions for detecting an EGMF in the range
$\simeq 10^{-10}-10^{-9}\G$. Since these
conditions are not unreasonable, UHE CRs have the potential
to provide important information on properties and origin of the
EGMF.

\acknowledgments
This work was supported by the DoE, NSF and NASA at the University of Chicago,
by the DoE and by NASA through grant NAG5-2788 at Fermilab, and by
Alexander-von-Humboldt Foundation. S.L. acknowledges the support of the POSCO
Scholarship Foundation in Korea.

\newpage
\begin{figure}
\plotone{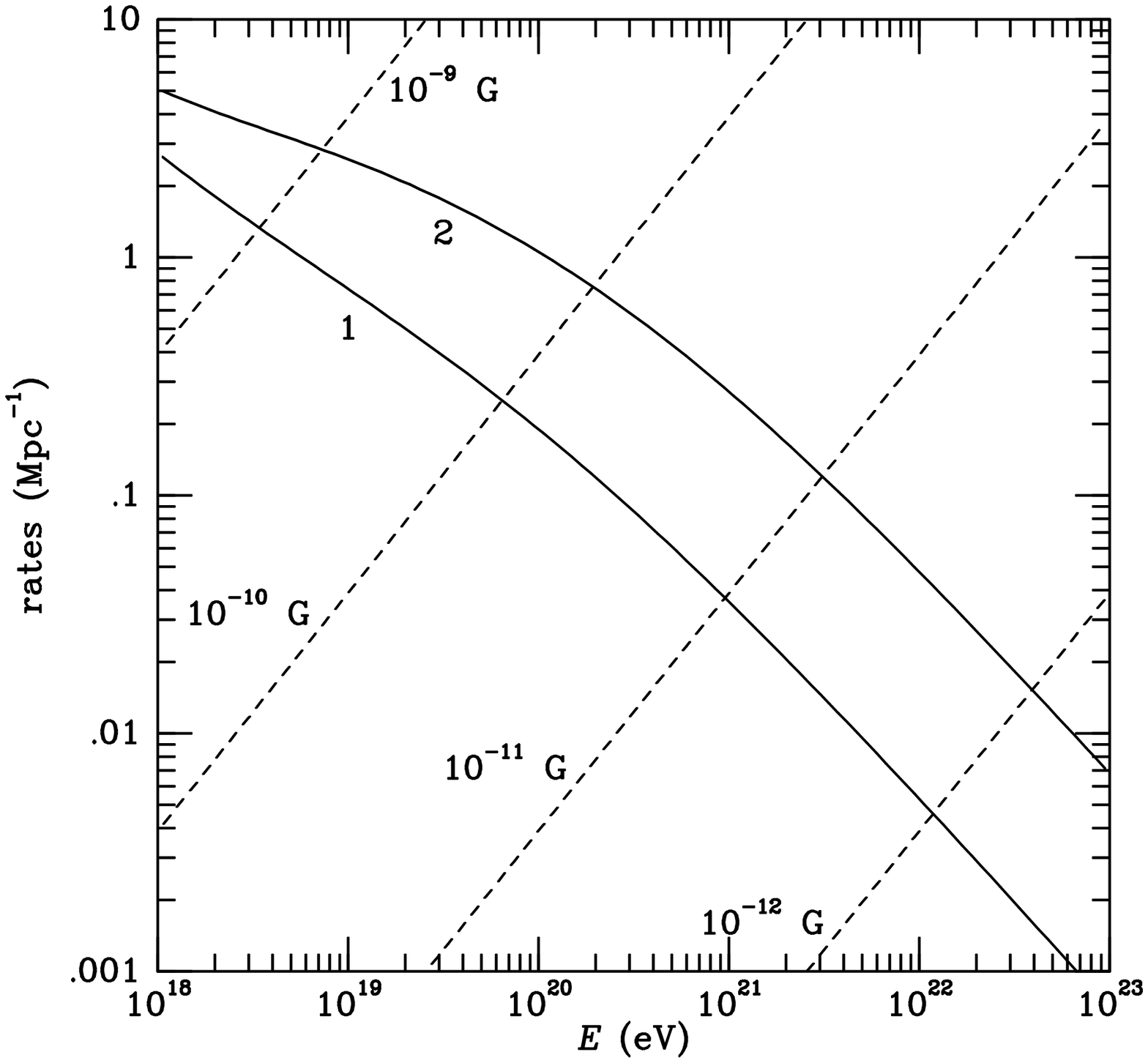}
\caption{The rate of inverse Compton scattering (ICS; solid
lines) and the fractional synchrotron loss rate (dashed
lines) for different magnetic fields as
functions of energy. 1 is for a frequency cutoff at 2 MHz in the diffuse
radio background, and 2 is for 0.5 MHz, respectively.}
\end{figure}
\begin{figure}
\plotone{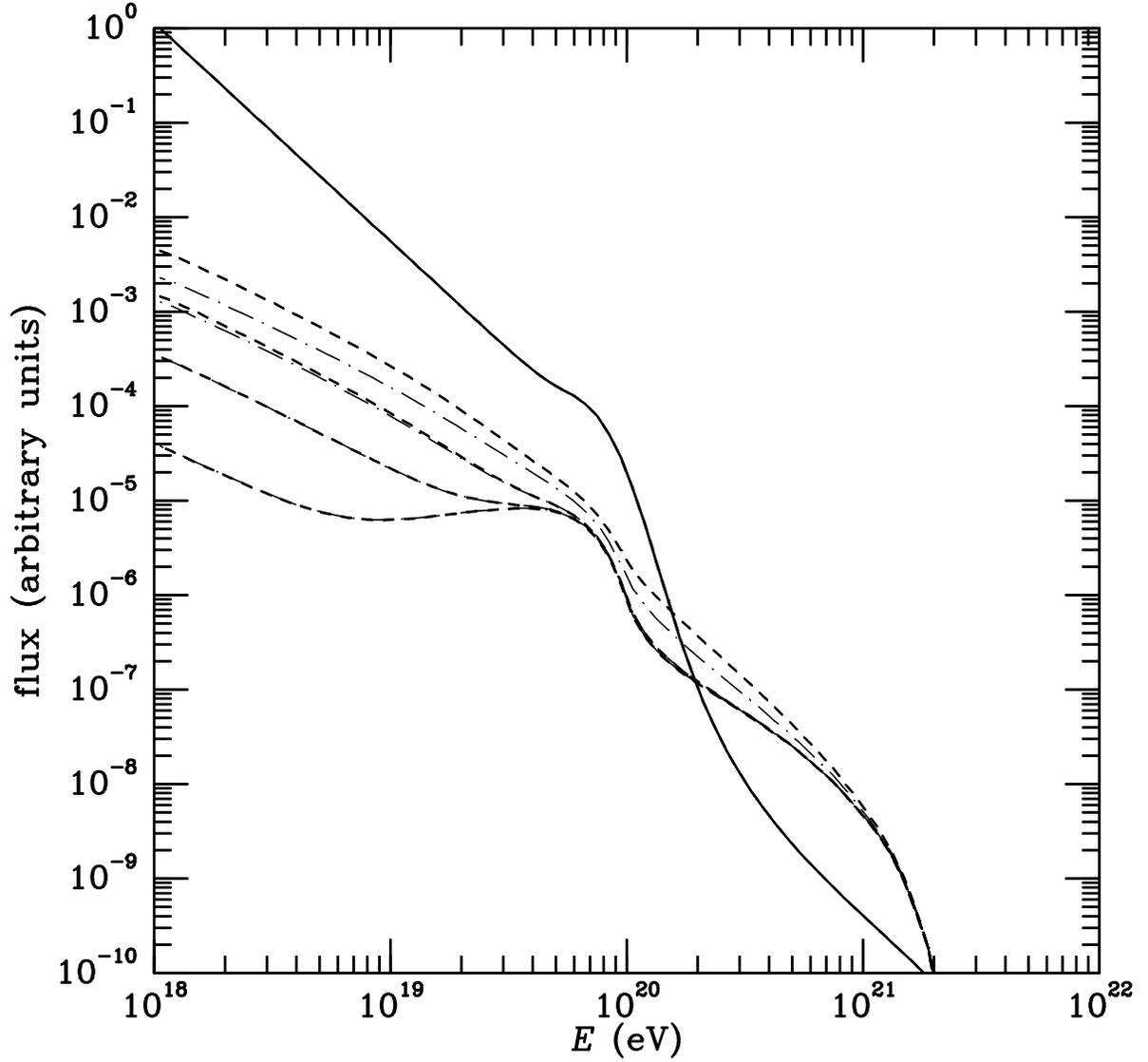}
\caption{Cosmic ray nucleon spectrum (solid line) from a single
source at a distance of $30\Mpc$, and secondary $\gamma$-ray
spectra for different
magnetic fields. The short dashed lines are with no intervening objects, and
the
long dashed lines with dots are with a 5 Mpc sized object located at 12.5 Mpc
from the
observer. The magnetic field values are $B=0, 10^{-10}, 3 \times
10^{-10}$, and $10^{-9}\G$, respectively from the top.}
\end{figure}
\begin{figure}
\plotone{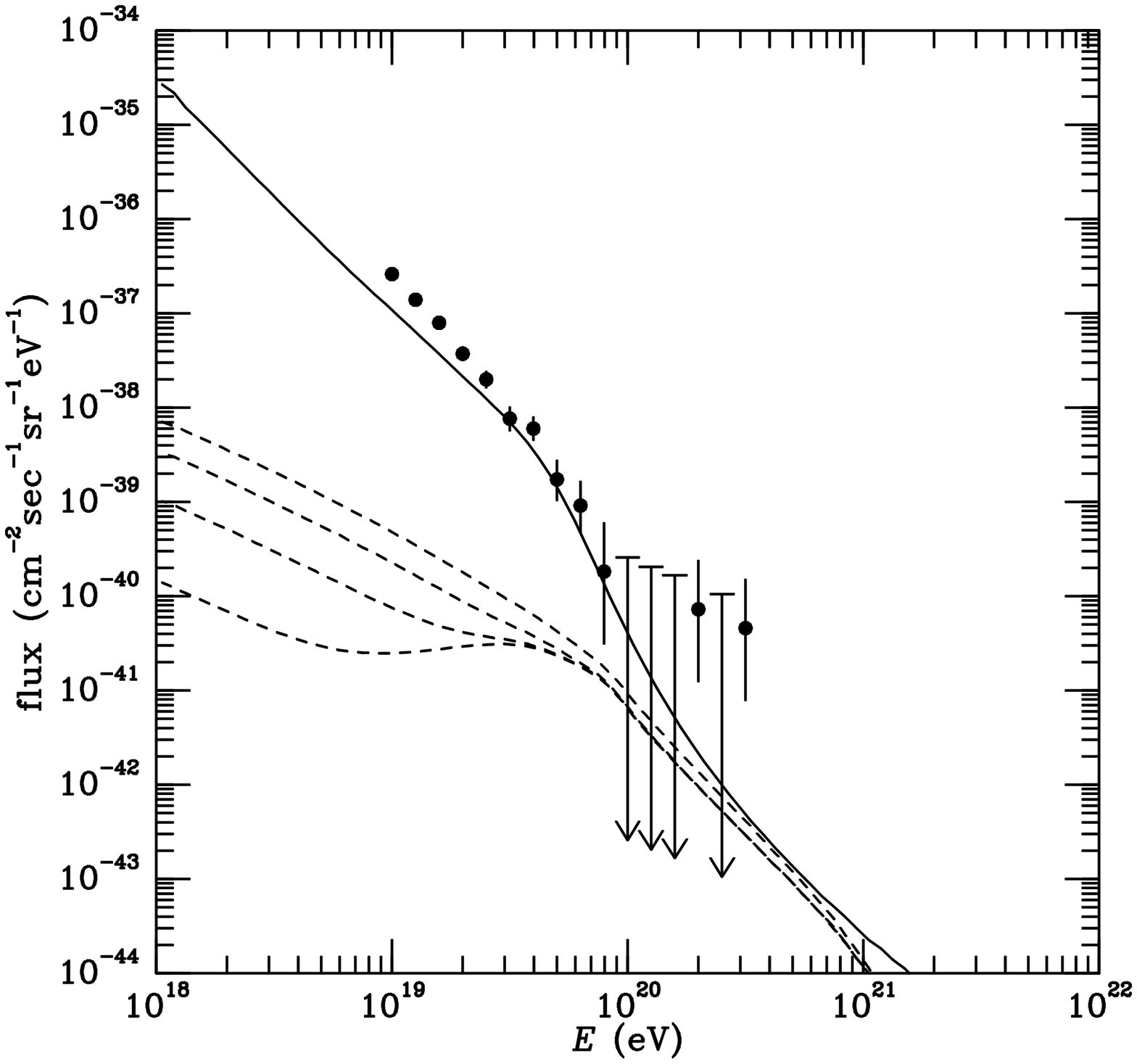}
\caption{Cosmic ray nucleon spectrum (solid line), and secondary
$\gamma$-ray spectra (dashed lines) for a continuous source
distribution up to 1Gpc distance whose comoving density scales
as $(1+z)^2$. The sum of the nucleon and \g-ray flux is fitted
to the combined data from the
monocular Fly's Eye (Bird et al.~1994) and the AGASA experiment
(Yoshida et al.~1995). Shown are $1\sigma$ error bars, and
84\% confidence level upper limits. The magnetic field values are
the same as in fig.~2.}
\end{figure}

\begin{references}
\reference{bgd} Berezinskii, V. S., Grigor'eva, S. I., Dogel', V. A. 1989,
JETP 69, 453.
\reference{bird} Bird, D. J., et al. 1994, \apj, 424, 491.
\reference{bora} Boratav, M., et al. eds. 1992, Nucl. Phys., 28B.
\reference{cba} Clark, T. A., Brown, L. W., Alexander, J. K. 1970, \nat,
228, 847.
\reference{grei} Greisen, K. 1966, \prl, 16, 748.
\reference{hs} Hill,  C. T., Schramm, D. N. 1985, \prd, 31, 564.
\reference{jko} Jedamzik, K., Katalinic, V., Olinto, A. V. 1995, in
preparation.
\reference{kkgv} Kim,  K-T., Kronberg, P. P., Giovannini, G., Venturi,
T. 1989, Nature, 341, 720.
\reference{kt} Kolb, E. W., Turner, M. S. 1990, The Early Universe
(Redwood City: Addison-Wesley).
\reference{kron} Kronberg, P. P. 1994, Rep. Prog. Phys., 57, 325.
\reference{lee} Lee, S. 1995, \prd, to be submitted.
\reference{ls} Lee, S., Sigl, G. 1995, in Proceedings of the 24th
International Cosmic Ray Conference, to be published (Rome)\notetoeditor{The
24th ICRC is to be held on August 28th, 1995, and the information about the
Proceedings such as the editors and the publishers is not available at the
time of submission}.
\reference{nbh} Nichol, R. C., Briel, U. G., Henry, J. P. 1994,
\mnras, 267, 771.
\reference{nma} Norman,  C. A., Melrose, D. B., Achterberg, A. 1995, \apj, in
press.
\reference{plag} Plaga, R. 1995, \nat, 374, 430.
\reference{rsb} Rachen, J. P., Stanev, T., Biermann, P. L. 1993, \aap, 273,
377.
\reference{slsb} Sigl, G., Lee, S., Schramm, D. N., Bhattacharjee, P. 1995,
FERMILAB-Pub-95/148-A, submitted to Science.
\reference{sdss} Stecker, F. W., Done, C., Salamon, M. H., Sommers, P. 1991,
\prl, 66, 2697.
\reference{sp} Szabo, A. P., Protheroe, R. J. 1994, Astropart. Phys., 2,
375.
\reference{ww} Wdowczyk, J., Wolfendale, A. W. 1990, \apj,
349, 35.
\reference{yt} Yoshida, S., Teshima, M. 1993, Prog. Theor. Phys., 89, 833.
\reference{yosh} Yoshida, S., et al. 1995, Astropart. Phys., 3, 105.
\reference{zk} Zatsepin, G. T., Kuz'min, V. A. 1966, Pis'ma
Zh. Eksp. Teor. Fiz., 4, 114 [JETP. Lett., 4, 78].
\end{references}
\end{document}